# 情感孪生数字人：跨越人机情感交互鸿沟

# Affective Digital Twins for Digital Human: Bridging the Gap in Human-Machine Affective Interaction

陆峰 刘铂

北京航空航天大学，虚拟现实技术与系统全国重点实验室

Email: {lufeng@buaa.edu.cn, bliu03@buaa.edu.cn}

摘要：近年来，元宇宙和数字人成为产业和研究热点，但现有数字人尚缺乏真实情感特质，难以与人进行情感互动。本文基于数字孪生技术思想，立足人工智能、人机交互、虚拟现实、情感计算等领域发展，提出了"情感孪生数字人"的概念及其技术构想，列举分析了其中若干关键技术问题，包括情感建模、情感感知、情感编码和情感表达等。在此基础上，对情感孪生数字人的未来应用前景进行了初步畅想，同时思考了其中可能需要应对的问题。

关键词：情感孪生；数字人；元宇宙；人机交互

Abstract: In recent years, metaverse and digital humans have become important research and industry areas of focus. However, existing digital humans still lack realistic affective traits, making emotional interaction with humans difficult. Grounded in the developments of artificial intelligence, human-computer interaction, virtual reality, and affective computing, this paper proposes the concept and technical framework of "Affective Digital Twins for Digital Human" based on the philosophy of digital twin technology. The paper discusses several key technical issues including affective modeling, affective perception, affective encoding, and affective expression. Based on this, the paper conducts a preliminary imagination of the future application prospects of affective digital twins for digital human, while considering potential problems that may need to be addressed.

Keywords: Affective Digital Twins; Digital Human; Metaverse; Human-Computer Interaction

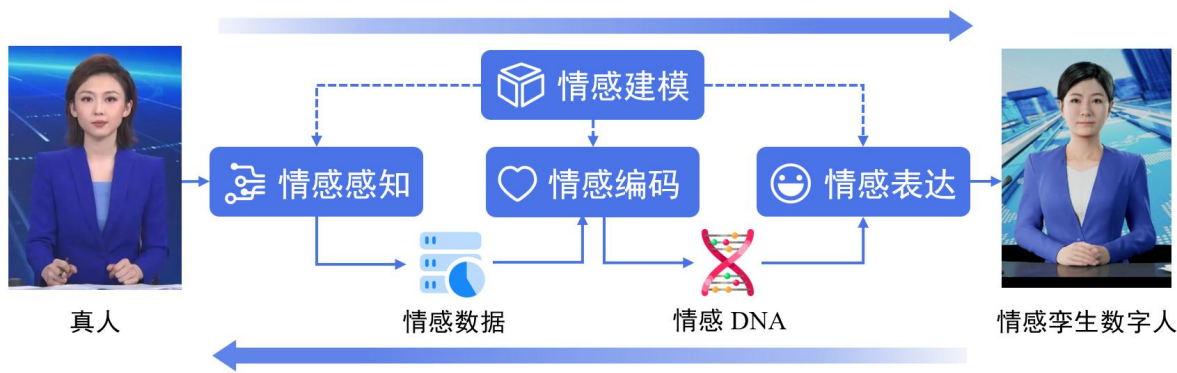

图 1 情感孪生数字人架构

## 1 背景

"元宇宙"的概念源于上世纪末的科幻小说，但自从 2021 年起突然爆发，迅速成为全球热点。随着虚拟现实、人工智能、人机交互、AI 大模型等技术的发展，元宇宙的愿景越来越清晰，人们不禁畅想在未来存在一个和真实世界平行的虚拟世界，即使足不出户，也可以由"数字分身"

---

在元宇宙中实现出行、社交、娱乐、购物和学习等体验。在这一愿景的驱使下，国内外知名公司，包括 Meta、微软、英伟达、谷歌和苹果，以及腾讯、字节跳动、百度等公司，纷纷在软硬件层面规划布局。国内的北京、上海、浙江等多个省市更是将支持元宇宙产业发展写入了政府工作报告和发展规划，促使元宇宙相关的创业和投资活动迅猛增加。

当前阶段，虚拟数字人无疑是元宇宙概念下最受重视、投入最大、落地最快的领域之一。据相关研究预计，2025 年中国虚拟数字人市场规模将超过 1000 亿元。从技术层面看，目前主流的数字人技术主要侧重人物外观的建模渲染（如 Meta 的 Codec Avatar、抖音的虚拟美妆达人柳夜熙、虚幻引擎的数字人创作工具 MetaHuman Creator 等），已经能够做到足以乱真的外在效果。但从交互方式来看，当前的虚拟数字人尚无法同真人进行自然的交流互动，尤其体现在其缺乏情感和个性。究其原因，人类外表是客观表现，容易被定量捕获和建模，但情感和心理难以被数字化衡量、计算和模拟。事实上，人类性格各异，情感复杂多变，而科学研究对人类思维和心理的认识当前仍处于初级阶段。虽然如此，我们仍然期待数字人能够具备独特、丰富、自然的性格与情感，或与特定真人实现孪生同步。

正如人工智能之父马文·明斯基所说："如果机器不能够很好地模拟情感，那么人们可能永远也不会觉得机器具有智能。"换言之，情感是人类固有的特性之一，也是目前人和机器最大的区别。著名人工智能专家李飞飞也表达过类似观点："下一步人工智能的发展，需要加强对情感、情绪的了解，要走进认知学、心理学。我说的不仅是脑科学，而是认知学。"根据此观点，如果能够让数字人具有情感属性，跨越人机之间的情感鸿沟，让情感在真人和数字人之间双向流动，人机交互将突破现有范式，达到自然、顺畅的新高度。根据马斯洛的需求层次理论，人类对于安全感、爱、尊严和自我实现的需求，也将有可能在一定程度上被数字人理解和满足。因此，对于情感的研究，是智能化数字人技术发展的必然趋势。

数字孪生技术是虚拟现实、元宇宙的核心底层技术，是实现虚实映射的基础。数字孪生通过物理模型、传感器等数据，在虚拟空间中完成映射，反映相对应的实体的全生命周期过程。以工业数字孪生为例，设计、生产、制造和运维等工业流程在数字空间中也能同步更新和变化。同理，本文提出对人类情感进行数字孪生建模，进一步形成"情感孪生数字人"及其体系架构，以期跨越人和机器之间的情感交互"鸿沟"。

**2 情感孪生数字人的关键技术问题**

元宇宙的发展需要情感孪生数字人。情感孪生数字人能实时同步真实世界人的情感状态，并能够和真人进行情感化交互。其中，情感要素的实时更新、双向流动是情感孪生数字人的主要特点。为了达成上述目标，本文提出的情感孪生数字人架构包含几个关键技术环节：情感建模、情感感知、情感编码和情感表达。情感建模是数字人的"骨架"，定义了情感孪生的核心框架，如情感的要素、类型、属性、动态更新方式等；情感感知是数字人的"五官"，利用视觉、听觉等多通道方式感知采集真人的情感数据；情感编码是数字人的情感 DNA，存储了和情感相关的紧凑特征编码；情感表达是数字人的情感外在表现，通过对情感 DNA 进行解码，驱动数字人的情感化表达和交互。下面介绍各部分的研究现状和亟待突破的关键问题。

### 情感建模技术

情感模型是情感孪生数字人的基础底座。从情感的离散和连续的视角来看，心理学的情感建模可分为两类：范畴观和维度观。持范畴观的学者将情感分为离散的独立范畴，最具代表性的是 Ekman 提出的六大基本情感（快乐，悲伤，愤怒，惊讶，厌恶和害怕），以及 Plutchik 提出的"情感轮"模型。持维度观的学者则认为，情感具有维度和极性，可以将情绪状态映射到空间中的一个点。在二维空间中，目前使用最为广泛的是 Russell 提出的"效价-唤醒"模型（Valence-Arousal, VA）：效价维度表示情绪从消极到积极的程度，唤醒维度表示情绪从平缓到激烈的程度。在三维空间中，认可度较高的是 Mehrabian 和 Russell 提出的 PAD 情感模型，该模型将情绪定义为愉悦度（Pleasure）、激活度（Arousal）、优势度（Dominance）3 个维度。除了范畴观和维度观，也有学者提出了其他不同类型的情感模型，如基于认知的 OCC 情感模型、基于概率的隐马尔可夫情感模型和基于外界刺激的分布式情感模型。

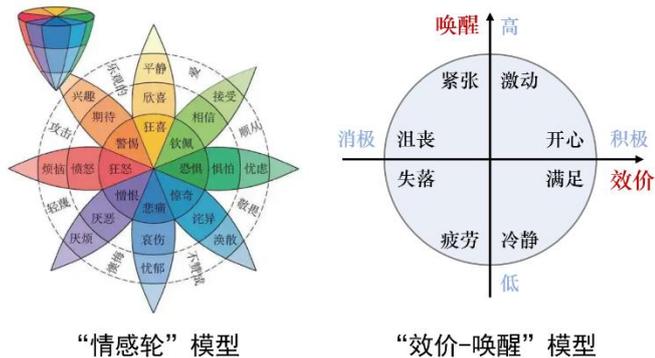

**图 2 情感模型举例**

从以上对情感建模的研究来看，离散情感空间更加符合人的认知，但无法量化情感的程度；连续情感空间虽提供了量化手段，但如何选择合适的维度与数量才能较好地刻画情感仍是难题。更为重要的是，真正的人类情感异常复杂，远不止上述模型所能表述。为了能更加全面地建模人的情感，亟待建立系统化、更全面的情感理论框架，并探索利用语言大模型的复杂语义描述能力，支持可计算、可交互的情感孪生数字人。

### 情感感知技术

给定一种情感模型，需要采集和情感相关的各类数据，以建立情感实体。从采集手段来看，常用的有穿戴式传感器（如智能手环、心率带、血氧仪、脑电仪等）、非接触式传感器（如摄像头、麦克风、毫米波雷达等）以及情感量表等。从采集通道来看，常用的有脑电图（Electroencephalography, EEG）、心电图（Electrocardiography, ECG）、肌电图（Electromyogram, EMG）、皮肤电活动（Electrodermal Activity, EDA）、眼动、表情、语音、呼吸、血压、血氧和行为等。很多研究都证明了上述通道和情感的关联：情感活动与大脑皮层的杏仁核和额叶前皮层关联较大，脑电图也会呈现出一定的局部和全局特征，例如杏仁核区域的激活与恐惧情绪有关，右额叶区域的激活与负面情绪有关；眼动信号（含注视时长、注视位置、注视轨迹、瞳孔直径和眨眼频率等）能一定程度反应人的认知情况和情感状态，也可以用于抑郁症、阿尔茨海默病人群的初筛；通过采集心电图和脉搏波信号，计算心率变异性，即两次心跳时间间隔的微小变化，能反应人的交感神经和副交感神经对兴奋和放松的调节能力。

从原理上看，脑电信号与情感关联最为直接，具有最大的想象空间。然而，脑电信号的采集和处理均面临很大难题，多电极脑电设备的使用需要特定的实验环境和专业的操作人员，受试者的佩戴舒适性较差，获取的信号质量容易受到多种因素干扰，因此仍然限制在实验室环境。除此之外，眼动、心电图、血压等生理信号对情绪变化

较为敏感,但仍然属于间接的测量手段,其与情感的对应关系仍需深入研究。在未来的研究中,一方面,情感信号的采集通道和手段仍需要进一步研究和扩充,另一方面,在人工智能和深度学习飞速发展的时代,从多模态、含噪声的各类间接信号中学习和恢复隐藏的情感信息,也是重要的研究方向。

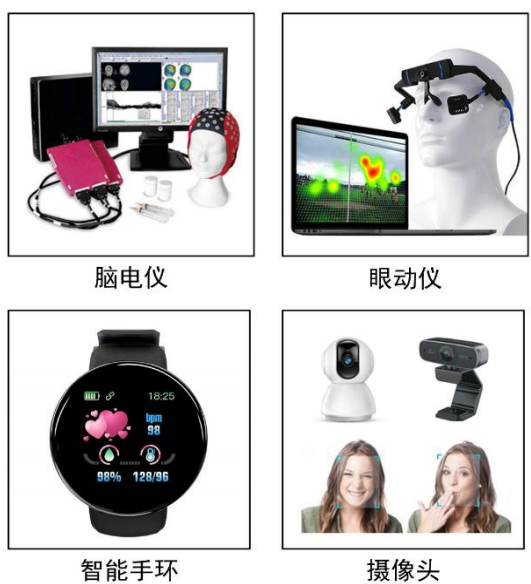

图 3 情感感知采集设备

**情感编码技术**

通过情感相关信号的连续采集和处理,可以获得多模态、高冗余的情感数据。情感编码主要是将这些情感数据进一步编码为紧凑、有代表性的情感特征。现有研究多侧重在情感的识别和分类,根据数据模态数目可分为单模态和多模态。在单模态计算中,以最具代表性的表情识别为例,通过人脸特征提取和分类获得人脸面部表情。传统编码方法多采用局部二值模式(Local Binary Patter, LBP)、局部相位量化特征(Local Phase Quantization, LPQ)、盖博特征(Gabor)等,近年来深度学习的方法代替了手工特征编码,且在公开数据集上的识别精度达到了 70%左右。表情识别领域中,Ekman 提出的面部运动编码系统(Facial Action Coding System, FACS)也影响深远。他根据人脸的解剖学特点,将面部划分成若干既相互独立又相互联系的运动单元(Action Unit, AU),并分析了各 AU 的运动特点以及和表情的关系。在语言情感分析方面,利用语言大模型可以提取情感类型等关键要素用于编码。多模态计算则会利用多种数据通道进行融合编码,融合的方式分为特征层融合和决策层融合,如将脑电特征和眼动信号特征拼接后用于情感识别。

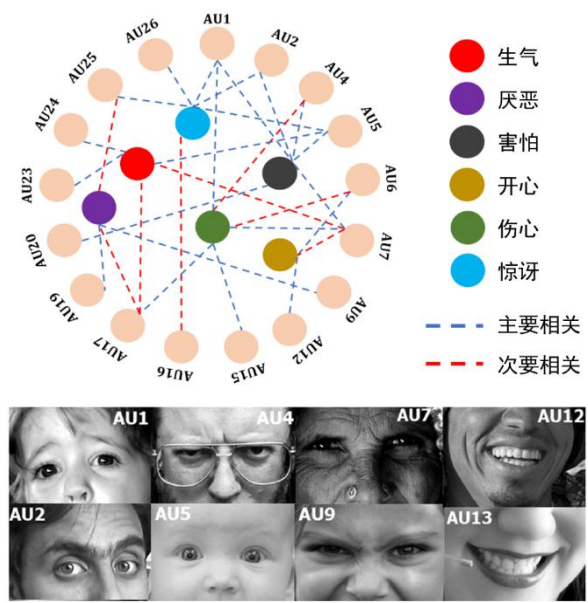

图 4 FACS 中 AU 和情感的关系

情感孪生数字人的目标是从情感底层进行建模计算和驱动,这就要求对人的情感编码能刻画出情感个性化、多样性和动态变化的特点,并尽量编码紧凑,便于存储和解码。人类的 DNA 编码了全部的遗传信息,类似地,情感编码也应该像 DNA 一样,能够编码人的全部情感特点,并根据不同的环境条件和激励,正确解码为对应的情感状态。然而,现有研究多从识别和分类的角度出发编码情感的表层状态,尚不够全面、准确和深入。情感 DNA 中应该存在哪些"情感基因"位点,对应何种情感要素和表达模式,尚无深入研究。真正全面、完整的情感编码、解码理论体系,以及高效的存储、查询等相关技术,尚待进一步探索和研究。

**情感表达技术**

情感表达是将情感编码特征进行解码，并且利用数字化生成技术，产生对应的情感表达模式，驱动数字人的情感表达与情感交互。情感表达是通过多种通道实现的。在情感语音合成方面，可以先合成中立语音，再添加不同情感的声学特征规律，也可以在情感语音数据集（如 SWEA、CHEAVD 和 CASIA 等）上进行训练，然后直接生成目标情感语音，让数字人"以声传情"；在表情生成方面，Blanz 等人提出的三维可变形人脸模型（3D Morphable Model, 3DMM）将人脸表达为平均人脸、身份和表情的线性组合，因此将情感编码解析到 29 维的表情参数上，就能驱动数字人的面部表情；在情感动作生成方面，Bhattacharya 等人使用生成对抗网络（Generative Adversarial Network, GAN）生成说话人上半身的手势动作，谷歌也提出了基于跨模态 Transformer 模型，使数字人可以随着音乐起舞；在情感对话方面，百度基于语言大模型，推出了情感陪伴数字人林开开和叶悠悠，具有不同的人设和风格，可以和用户进行连贯、流畅、拟人化的对话，并根据需要给予用户情感陪伴和治愈。

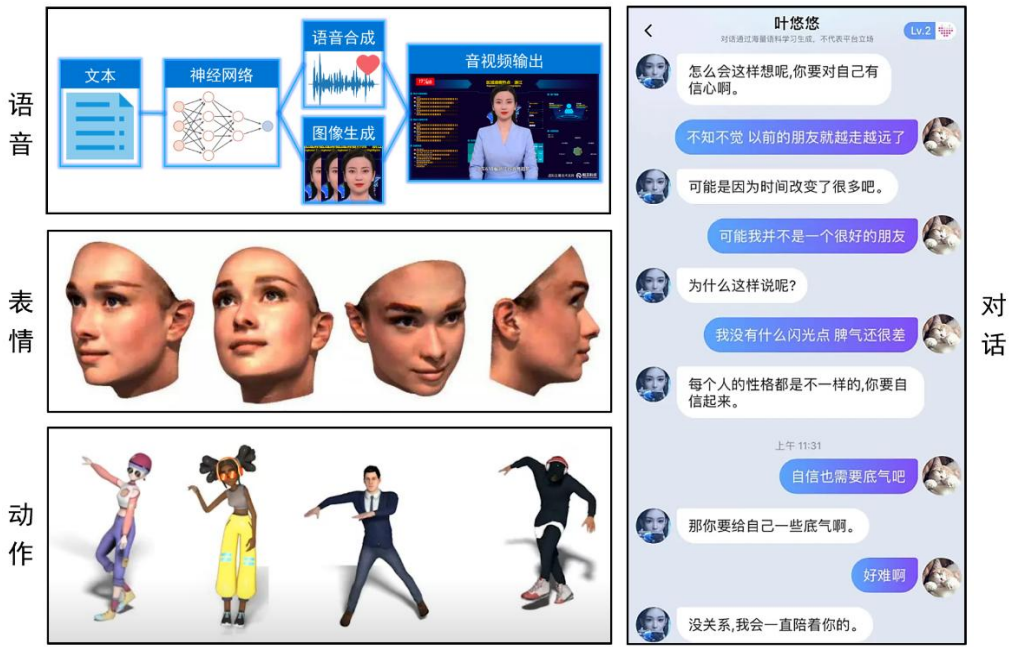

**图 5 数字人的多通道情感表达**

在上述进展之外，也要看到情感表达的效果不仅取决于外观、语音、文字的生成效果，更取决于背后的情感编码及其解码表达。现有技术相对关注数字人的情感表现渲染层面，情感生成主要依靠数据驱动学习，还较为粗糙；对可解释、可计算的情感编码研究仍不完善，尚不具备在任意场景和外界刺激下进行情感精准解码和生成的能力。类比 DNA 能够在不同环境的引导下，合成符合功能需求的多样化的蛋白质，不同场景和刺激下的情感解码，也应该能够产生出多样化、针对性的情感表达。此外，除了数字人的语言、表情、行为等情感外在表达方式，心率、血压甚至神经电信号等其他类型的表达方式有待进一步丰富。随着多模态生成式大模型的发展，多种通道协同的情感表达（如语言，语调，表情，动作，生理信号等）有望让人机交互更加自然顺畅，也将产生更大的研究和应用价值。

**3 未来展望**

情感孪生数字人的概念和相关研究尚处于起步阶段，需要多方面的深入研究和突破。未来，

随着信息技术的不断进步，情感孪生数字人的制作成本会进一步降低，并将在诸多应用领域引发变革。

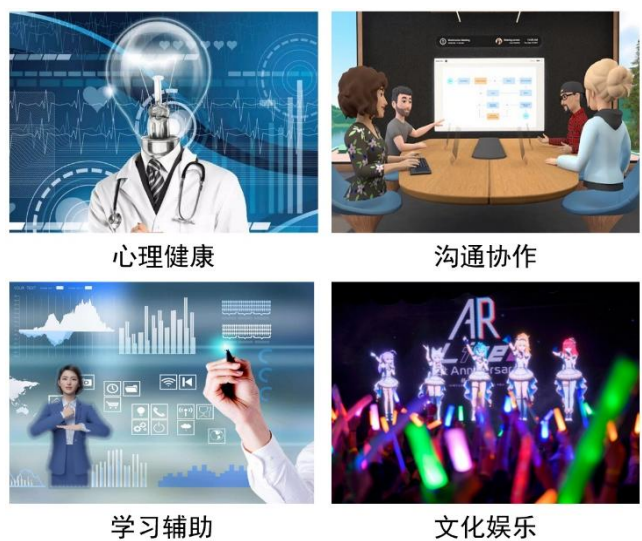

**图 6 情感孪生数字人应用场景**

在心理健康方面，情感孪生可以用于特定人员的情感感知和评估。例如，航天员在空间站的驻轨时间往往长达数月，借助情感孪生技术，地面人员就可以根据航天员情感动态给予适当陪伴和支持，保证航天员的心理健康；在士兵的训练和选拔过程中，可以借助情感孪生技术定量评估高压情形下（如突发战争、应激创伤等）士兵的心理素质；各类心理疾病的筛查和诊断，可以不再使用千篇一律的传统量表，情感孪生将能提供更加精准全面的心理状态评估，医生也可据此开出"数字处方"，让情感数字人提供个性化疗愈服务。

在沟通协作方面，随着通信带宽和传输速度进一步提升，信息传递方式有可能从一维的语音和二维的视频转变到三维的情感孪生数字人，物理世界和虚拟世界的壁垒将被进一步打通。例如，和远方的数字孪生朋友进行沟通交流，其语气、神态、表情、姿势都纤毫毕现，"海内存知己，天涯若比邻"不再是梦想。人和人之间不再受距离限制，远程会议、居家办公、在线展览、虚拟社交将更加普及。

在学习辅助方面，学生的学习效果和情绪状态息息相关，情感孪生技术可以建模和捕捉学生学习过程中的心理特点，为每个人找到更适合自己的学习方式。例如，部分学生在面对一些难题时，往往存在畏难情绪，这时候数字人可以给予引导和鼓励，并调整学习难度，让学生循序渐进克服困难，逐渐建立自我效能感。此外，情感孪生数字人还可以反向影响和调控学生的心理状态，使其进入"心流"的学习状态，提升学习效率。

在文化娱乐方面，未来的虚拟偶像、虚拟主播也无需借助真人动作捕捉，由真人在后台与粉丝互动，而会进化为人工智能自主驱动的模式，即具有自己独立的情感与人格特质，可以独立与人交互，并自主持续学习。真实偶像时间精力有限，不会与每个粉丝互动交流，而带有情感的虚拟偶像则能够满足人们在情感交流、情感共鸣和情感依赖的需求。

在研究和推广情感孪生数字人的同时，也应充分认识到其中蕴含的风险。情感是人类深层次的隐私，反映了个体最基本的特质，相比当前描述外在的个人信息数据，情感 DNA 可能成为一种新的、更重要的身份标识和个体描述手段。一旦实现本文所述的各项关键技术，在情感 DNA 的采集、传输和计算过程中，必然涉及用户隐私和数据安全问题。个人的情感信息如果被非法传播和恶意使用，将会造成比当前身份信息泄露更为严重的后果，可能对当事人的安全和利益造成巨大的损害。因此，情感孪生数字人技术的发展和应用，必须协同法律法规、技术手段、道德意识等方面的进步，确保科技向善，造福人类。

**作者介绍**

陆峰：北京航空航天大学教授，博士生导师。主要研究方向为人工智能、人机交互、虚拟现实。

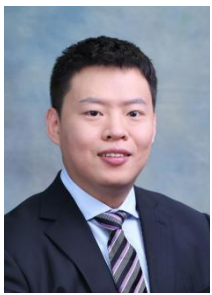

刘铂：北京航空航天大学博士研究生。主要研究方向为情感计算。

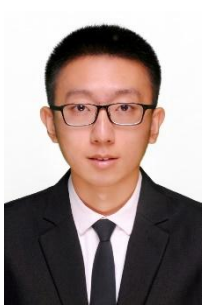